# Quiescence near the X-point of MAST measured by high speed visible imaging

N. R. Walkden[1], J. Harrison[1], S. A. Silburn[1], T. Farley[1,2], S.S. Henderson[3], A. Kirk[1], F. Militello[1], A.Thornton[1] and the MAST Team[1]

EUROfusion Consortium, JET, Culham Science Centre, Abingdon, OX14 3DB, UK

[1] CCFE, Culham Science Centre, Abingdon, OX14 3DB, UK
[2] Department of Electrical Engineering and Electronics, Univ. Liverpool, L69 3GJ, UK
[3] Department of Physics SUPA, University of Strathclyde, Glasgow, G4 0NG, UK
Email: `nick.walkden@ukaea.uk`

**Abstract.** Using high speed imaging of the divertor volume, the region close to the X-point in MAST is shown to be quiescent. This is confirmed by three different analysis techniques and the quiescent X-point region (QXR) spans from the separatrix to the $\psi_N = 1.02$ flux surface. Local reductions to the atomic density and effects associated with the camera viewing geometry are ruled out as causes of the QXR, leaving quiescence in the local plasma conditions as being the most likely cause. The QXR is found to be ubiquitous across a significant operational space in MAST including L-mode and H-mode discharges across maximal ranges of $9.8 \times 10^{19} m^{-2}$ in line integrated density, 0.36MA in plasma current, 0.11T in toroidal magnetic field and 3.2MW in NBI power. When mapped to the divertor target the QXR occupies approximately an e-folding length of the heat-flux profile, containing $\sim 60\%$ of the total heat flux to the target, and also shows a tendency towards higher frequency shorter lived fluctuations in the ion-saturation current. This is consistent with short-lived divertor localised filamentary structures observed further down the outer divertor leg in the camera images, and suggests a complex multi-region picture of filamentary transport in the divertor.

## 1. Introduction

Succesful exploitation of ITER requires careful control of the power to the divertor [1, 2] and understanding the tokamak exhaust is recognised as a key requirement for DEMO [3]. Excess heat fluxes to divertor surfaces will lead to degradation of the plasma-facing components and ultimately limit the operational lifetime of the machine. Several synergistic techniques to mitigate such machine limiting effects exist which are sensitive to particle and heat fluxes arriving in the divertor and impinging material surfaces. Predictions of these fluxes in next-stage tokamaks such as ITER and DEMO are hindered, however, by an incomplete understanding of cross-field transport processes in the plasma edge and scrape-off layer (SOL).

In the upstream SOL, above the X-point surrounding the plasma core, cross-field transport is robustly non-diffusive [4, 5]. Particles and heat are carried away from the core plasma through the SOL in intermittent coherent bundles known as filaments (alternatively blobs) which propagate through the SOL [6]. Two-dimensional transport codes, which make a diffusive approximation for cross-field transport, routinely require transport coefficients far in excess of classical values indicating the important role of filamentary and turbulent motion to both heat and particle transport. Filaments are thought to be important contributing factors to a number of phenomena in the upstream SOL, for example density shoulder formation [7], with recently developed models capable of interpreting SOL profile formation based purely on filamentary transport [8, 9]. Furthermore significant progress has been made towards understanding the motion of filaments with recent comparisons between state-of-the-art simulations and experimental measurements on TORPEX [10] and MAST [11] producing favourable results.

Transference of our understanding of transport based on filamentary physics to the divertor volume is complicated by the presence of the X-point. Strong magnetic shear and flux-expansion exist around the X-point as the poloidal magnetic field tends towards a null. These effects can significantly deform the shape of flux-tubes passing by the X-point [12] which has been predicted to impact the dynamics of filaments [13, 14, 15, 16, 17]. Such a deformation in the shape of filaments was observed on Alcator-Cmod [18] (although slightly above the X-point) and more recently on MAST [19, 20] in full views of the X-point region. Indirect evidence of X-point decorrelation of filaments was demonstrated on NSTX [21], though a direct observation has yet to be obtained. Also demonstrated on MAST was the existence of two classes of fluctuating filamentary structures localised to the divertor volume. These exist in the private-flux region (PFR) of the inner divertor leg and in the near-separatrix region of the outer divertor leg. Recent measurements on Alcator-Cmod have also observed these divertor localised filaments, though the details of their propagation differ [22]. This paper seeks to extend the characterisation of filaments in the divertor of MAST by focussing on the role of the X-point. In particular, by carefully interpreting high speed imaging of the divertor volume the X-point region will be shown to be quiescent in MAST.

The paper is organised as follows: Section 2 presents details of the imaging diagnostic used for this study. Section 3 characterises the nature of the X-point region in MAST. Section 4 extends the analysis to a database of MAST plasmas. Section 5 compares the result at the X-point with data taken at the divertor target of MAST and attempts to reconcile the two measurements. Section 6 discusses the implications of the quiescent X-point region and compares with other examples found in literature before section 7 concludes.

## 2. Diagnostic Details

The data analysed within this paper was recorded on a photron SA1.1 camera‡ with a frame-rate of 120kHz, an integration time ranging from 5.6 to 8.3$\mu$s and a readout window of 160x192 (horizontal x vertical) pixels on the detector providing $\sim$ 5mm spatial resolution in the poloidal plane at the camera tangency angle. The camera is unfiltered and sensitive to the visible spectrum. As such it is dominated by Balmer-$\alpha$ emission in the deuterium plasmas observed. The camera view is directed tangentially into the MAST vessel and encompasses a poloidal plasma cross-section extended from approximately 30cm above the lower X-point to the strike point on the divertor target. Figure 1 illustrates the poloidal view of the camera, alongside a false color image from the camera overlaid on top of a CAD visualisation of the MAST vessel.

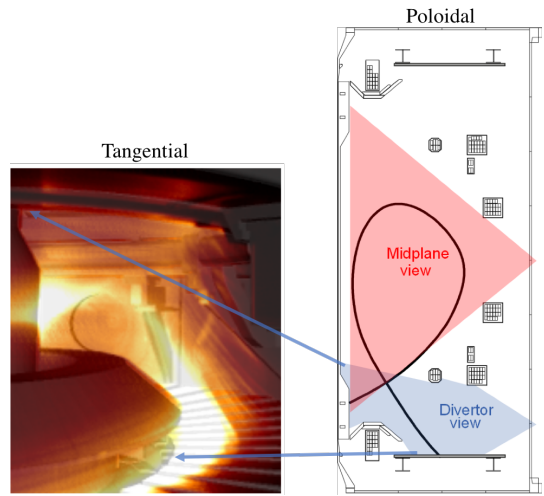

**Figure 1.** False color image taken from the SA1.1 fast camera tangential view of the MAST divertor overlaid onto a CAD visualisation of the MAST vessel. Also shown is a poloidal projection of the camera view with the divertor view (as well as the standard midplane view) of the camera highlighted.

Real space coordinates are mapped onto the camera view by registering features observed in the camera image to their real-space locations on a CAD model of the vacuum vessel using the `calcam` code§. For a significant portion of the analysis conducted here a background subtraction has been applied to the movies to isolate the fluctuating component of the emission picked up by the camera from the steady state emission. This is achieved by determining a 'background' from the pixel-wise minimum of the current movie frame and the 19 preceding frames. This background is subtracted from the current movie frame to produce an image of the fluctuations in the frame. This method has been used successfully for the analysis of filaments in the main chamber of MAST [23, 24] and in the divertor [20]. Alternative methods for background subtraction were compared, including a background formed from the pixelwise mean and median of the history of frames alongside a Fourier based filter. The pixelwise minimum based background subtractor was found to perform best with regards to extracting the fluctuating structures of the movie, though no results presented in this paper changed when the background subtractor was altered. Figure 2 illustrates the process of background subtraction by presenting images of a raw frame, the fluctuating component of the image and the background.

‡ http://photron.com/high-speed/cameras/fastcam-sa1-1/
§ Available at https://github.com/euratom-software/calcam

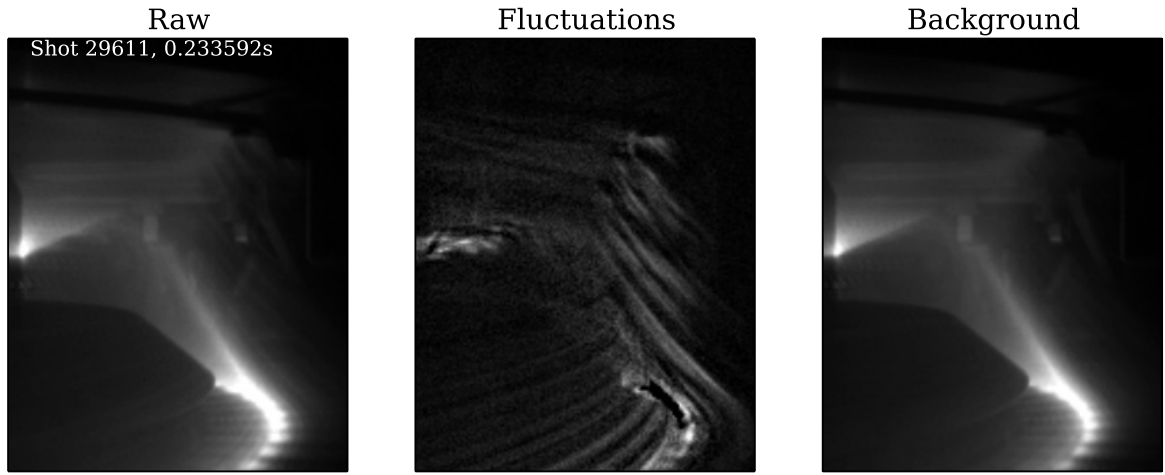

**Figure 2.** Raw (left), fluctuating (center) and background (right) images from MAST shot 29611 demonstrating the use of background subtraction to isolate fluctuations in the movie.

## 3. Characterisation of the X-point Region

*3.1. Experimental observations*

This section focusses on the analysis of the MAST plasma from shot 29611, which is a lower single-null Ohmic L-mode with a 600kA plasma current (see table 1), during a time period from 0.23335s to 0.25175s using the camera setup described in the previous section. In particular an emphasis is placed on the region close to the X-point (the position of which is inferred from EFIT [25]). To characterise general properties of the signal measured by the camera pixel-wise statistical moments are taken (statistical moments of the time-series of each individual pixel in the camera). The distribution of these moments on the image plane of the camera then provides insight into the fluctuating nature of the movie. The statistical moments are calculated both for the raw movie data and for the background subtracted data, providing a quantification of the effect of the background subtraction. Figure 3 shows the first three statistical moments (mean, standard deviation and skewness).
Comparing the results of the statistical analysis from figure 3 between the raw and background subtracted versions of the movie shows that only the first moment (the mean) is significantly impacted by the background subtraction. In the raw movie case, the mean value rises radially towards the X-point, whilst in the background subtracted data the converse is true. In the standard deviation and skewness some variation is apparent between the raw and subtracted data, however the structure remains consistent. This shows that the characteristics of the fluctuations in the movies are not altered by the process of background subtraction as expected.

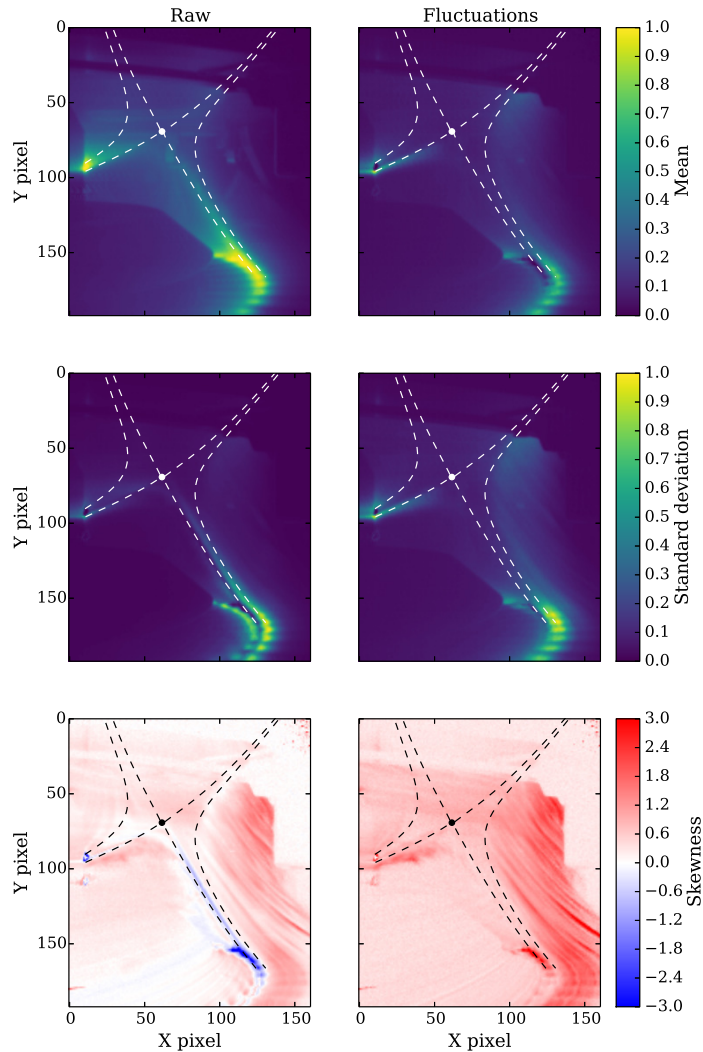

**Figure 3.** Pixel-wise mean (upper row), standard deviation (central row) and skewness (lower row) for the raw movie (left column) and the background subtracted movie (right column). The separatrix and $\psi_N = 1.02$ poloidal flux surface at the camera tangency angle have been overlaid as dashed lines, whilst the position of the X-point is shown with a filled circle.

Close to the X-point, the standard deviation of the signal drops significantly. This is accompanied by a drop in the mean of the background subtracted movie, which suggests that the signal arriving at the camera from fluctuating quantities in the region of the image close to the X-point is decreased. The skewness of the signal displays a more complex pattern. Away from the X-point the skewness increases, but is also organised by structures that angle diagonally from upper left to lower right. Such structures correspond well to the filamentary structures shown in the background subtracted movie frame in figure 2. These structures are consistent with the poloidal cross-sections of filaments that originate upstream with approximately circular cross-sections, but are deformed along the magnetic fieldline and become highly elliptical [12]. Close to the X-point however this structure disappears and the skewness becomes homogenous. The non-zero nature of the skewness in this region close to the X-point can be attributed to filaments existing at larger radii poloidally, but wrapping around toroidally and passing in front and behind the poloidal plane. However the change in nature of the skewness close the X-point, along with the reduction in the standard deviation suggests that, in the poloidal plane, the plasma appears quiescent close to the X-point. The approximate point at which the plasma becomes quiescent conforms relatively well to a poloidal flux surface, here found to be the $\psi_N = 1.02$ flux surface. This bounding region between the separatrix and the $\psi_N = 1.02$ surface is highlighted in figure 3. This region will be termed here the Quiescent X-point Region (QXR). The QXR is somewhat localised poloidally with higher frequency filamentary that occupying the region emerging approximately half way along the divertor leg, towards the target. This will be highlighted more in section 5.

The changing nature of fluctuations observed on the camera in the QXR can be further elucidated by measuring the time-series of the intensity on camera pixels along a line-of-interest (LOI) extending from the X-point radially outwards. This is shown in figure 4 for a series of 300 frames taken within the time-period that the statistical moment calculations in figure 3 were measured.

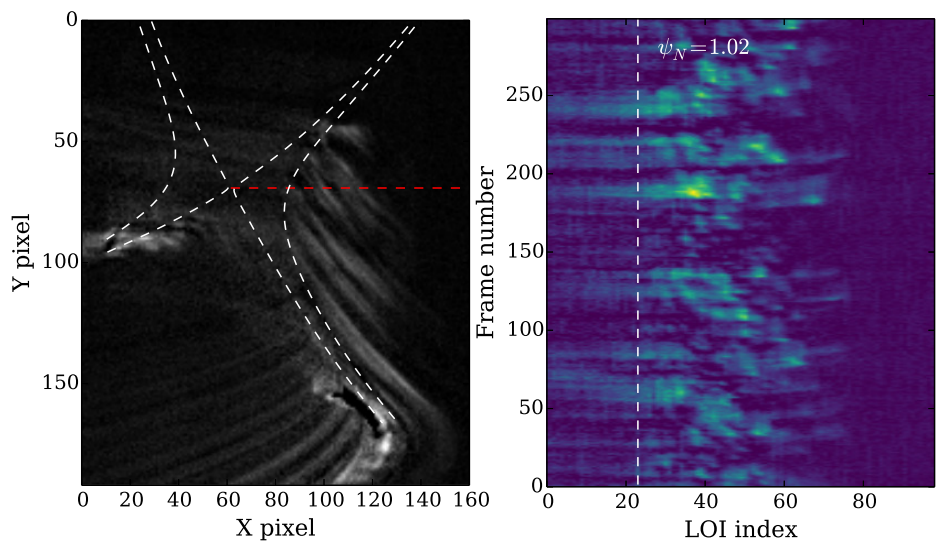

**Figure 4.** Left: Position of the LOI, which extends horizontally outwards from the X-point. Right: Time-series of the background subtracted emission measured along the LOI with the index at which the $\psi_N = 1.02$ crossing occurs shown.

Fluctuations measured on the LOI appear as strong bursts localised to a few pixels along the LOI outside of the $\psi_N = 1.02$ surface. Inside this surface the nature of the fluctuations changes with the entirety of the LOI becoming lit up each time a fluctuation occurs. This is consistent with the interpretation given in the analysis of the statistical moments of the movie. Outside the $\psi_N = 1.02$ surface the light emission is primarily from the poloidal cross-section of intermittent structures which cut across the LOI, producing the bursty nature that is observed in figure 4. Inside the $\psi_N = 1.02$ surface the emission is the result of filaments existing outside the QXR wrapping around in front and behind the poloidal plane and so the entirety of the LOI is lit each time a filament is observed and the parallel structure of the filament is sampled, rather than the perpendicular structure. The loss of fluctuations towards the far right of the LOI is due to shadowing by the P3 poloidal magnetic field coil.

As a third and final demonstration of the lack of observable filamentary structures within the QXR, cross-correlation analysis has been conducted on the same series of frames that contributed to figure 3. This involves correlating the time-series from a selected pixel with all other pixels on the camera sensor at zero time delay. Through this process the typical structure of a fluctuation on the camera image that is observed on the selected pixel is determined. Scanning the selected pixel radially outwards along the LOI from figure 4 shows how the structure of a typical fluctuation varies at different positions moving away from the X-point. This is shown in figure 5.

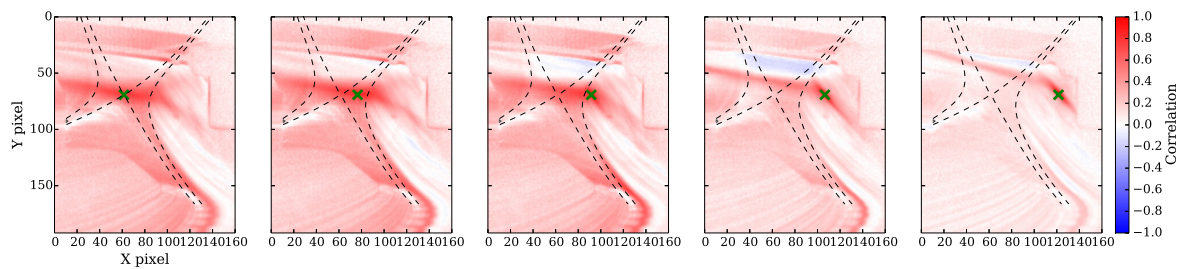

**Figure 5.** Cross-correlation of the target pixel (green cross) time-series with all other pixels in the camera frame, for five target pixel positions along the LOI of figure 4

Inside the QXR, for both positions of the selected pixel, the structure of the cross-correlation is a band extending mainly horizontally across the image but with no horizontal localisation close to the selected pixel, indicating once again that these fluctuations are a result of filaments passing behind the plasma and the structure observed on the camera is representative of the parallel structure of the filament. As the selected pixel moves outwards the region of heightened correlation changes shape becoming angled diagonally. The cross-correlation becomes highly localised in both vertical and horizontal directions and is now representative of the poloidal cross-section of a filament. There is a weak anti-correlation present above the filament band structure, however this is obscured by the P3 magnetic field coil and may be a measurement artifact. Interestingly outside of the QXR the filament cross-sections also correlate well with a region at the divertor target which moves outwards radially as the selected pixel is scanned outwards. This is present for selected pixels both in and outside the QXR, however for the inner pixels the filamentary structure that emerges has its cross-section outside the QXR in the poloidal plane and therefore correlates with a region on the divertor plate that is correspongingly outside the QXR. This shows that when filaments are present in the divertor, they connect all the way down to the divertor target. This filamentary contribution to divertor conditions has recently been investigated by Thornton *et al* on MAST in the context of the heat-flux to the divertor target [26].

The three measurements presented in this section are entirely consistent with one another and all indicate that, as observed on the camera, the light emission in the region between the X-point and (approximately) the $\psi_N = 1.02$ flux surface is quiescent in the poloidal plane. There are several complicating factors that may affect the camera measurement however, which must be ruled out before firm conclusions on the nature of fluctuations near the X-point can be established. In the next section these factors will be isolated and investigated in turn.

*3.2. Cause of the QXR*

The intensity measured on a pixel of the camera sensor is the result of a line integration of the plasma emissivity (dominated by $D_\alpha$) along a sightline. This emissivity depends upon the local neutral deuterium population, the electron density, and the photon emission coefficient which varies non-linearly as a function of $T_e$ and $n_e$. It is important now to establish whether the QXR is a result of: a) quiescence in the plasma conditions near the X-point; b) a screening of neutral density near the X-point or c) effects associated with the viewing geometry of the camera and its interaction with the 3D structure of the filaments in the divertor that affect the line-integration of the emission. Turning attention first to the role of the neutral density, screening of neutrals through, for example, dominant ionization in the outer SOL could produce the appearance of quiesence in the region near the X-point since there would exist no neutrals to become electronically excited and emit line radiation. Experimentally the neutral density along the divertor legs cannot be directly measured. The presence of visible radiation is a good indication of the presence of neutrals however, and in figure 2 emission can be tracked in up the divertor leg in the vicinity of the X-point in the background image. This emission pattern is similar to that of SOLPS simulations conducted by Havlickova et al [27], where it was shown that the contribution from Carbon radiation in MAST is localised to the target such that the emission in the divertor leg is primarily from neutral Deuterium. This indicates that neutral screening is not occuring and is not responsible for the presence of the QXR.

The role of the camera viewing geometry can be assessed through use of a forward model of the camera image via a synthetic diagnostic, details of which are given in the appendix. The magnetic geometry close to the X-point has a severely deformative effect on the cross-section of filaments [12, 13, 14, 16, 17] with filaments that are approximately circular at the outboard midplane becoming stretched by flux-expansion and sheared through the magnetic shear as they pass by the X-point. The result is that filaments in the divertor appear as thin ribbon-like structures with a disparity in their spatial dimensions in the cross-field plane. This structure can be seen clearly in figures 2 and 5. Furthermore the reduction in the poloidal field means that filaments distributed in the full 360 degree toroidal annulus at the midplane occupy a relatively compact region poloidally near the X-point. Figure 6, taken from ref [17], gives an illustration of the degree to which a filament cross-section may become sheared in the divertor.

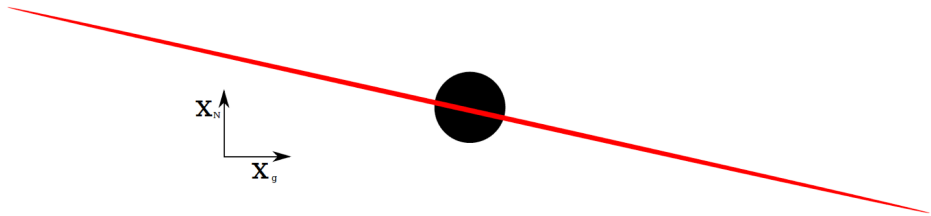

**Figure 6.** Illustration of the shearing induced in a filament cross section in the divertor (red) compared to its upstream circular shape (black). The directions normal ($x_N$) and binormal ($x_g$) are labelled. The shearing applied is a to-scale reresentation of that in MAST.

It is plausible that the reduction in filament spatial scale due to shearing and compacting in the poloidal plane could lead to an inability of the camera at its present resolution to distinguish individual filaments near the X-point. To test this an ensemble of synthetic filaments has been generated with positions, Gaussian widths and amplitudes sampled from experimentally relevant statistical distributions. In particular, filaments are seeded with a toroidal position at the outboard midplane sampled from a uniform distribution. Both the radial and toroidal widths of the filaments are sampled from log-normal distributions whilst the amplitudes are sampled from an exponential distribution. The radial position is sampled from a log-normal distribution with a minimum possible radial position specified by the parameter $R_{min}$. The choice of statistical distributions used here is based off experimental measurements on MAST [28], NSTX [29] and Alcator C-Mod [30]. An image of the ensemble of filaments as they appear in the divertor is then generated using the camera geometry of the experimental case, examples of which are presented in the appendix to this paper. 1000 images are produced, each representing a random realisation of the statistical distributions mentioned above, and treated in the same manner as a movie. The analysis techniques already used to characterise the QXR in MAST can now be applied directly to synthetic data which has known properties. In particular the cross-correlation analysis has been carried out for two sets of synthetic data. In the data set a) $R_{min} = 1.35$m such that filament positions upstream span the separatrix, whilst in data set b) $R_{min} = 1.367$m which corresponds roughly to the radial position of the $\psi_N = 1.02$ flux-surface at the midplane, so that there are no filaments with centers inside the QXR. In figure 7 the cross-correlation along the LOI from figure 4 is shown for these two synthetic datasets.

Outside of the QXR the cross-correlation shows a very similar structure in both synthetic data-sets as well as in the experimental data set shown in figure 5. This suggests that the interpretation of the filamentary structures in the outer SOL as filaments from upstream connected through to the target is robust. Inside the QXR the structure of the cross-correlation differs between data-sets a) and b). In data-set a), which contains filaments with positions spanning the separatrix, the cross-correlation inside the QXR shows a highly sheared but unmistakably elliptical filamentary structure. This structure is not present in data-set b) where no filaments are seeded inside the QXR. This shows that, if filaments exist within the QXR then the camera setup and analysis methods used here are sufficient to observe them. In addition to the results shown in this section, synthetic data-sets have been produced to investigate the effect of a finite camera integration time, a finite filament velocity and varying levels of noise. None of these aspects had any bearing on the results and cannot be held responsible for the QXR.

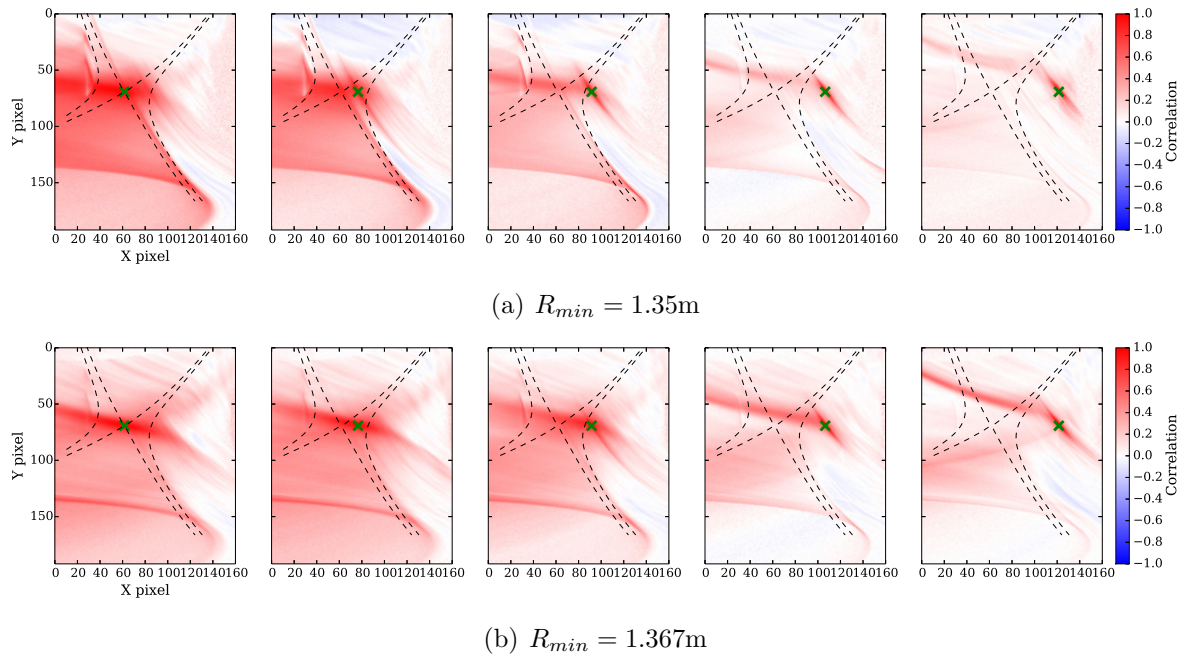

(a) $R_{min} = 1.35$m

(b) $R_{min} = 1.367$m

**Figure 7.** Cross-correlation analysis carried out in exactly the same manner as in figure 5, but analysing two synthetic datasets, a and b. Importantly filamentary structures can be identified within the X-point region in dataset a, but not in dataset b.

In this section neutral screening effects and camera geometry effects have been eliminated as causes of the QXR, leading to the conclusion that the quiescence observed represents quiescence in the local plasma conditions in the poloidal plane within the QXR.

## 4. QXR presence in MAST plasmas

Having established that the QXR observed in MAST is a result of quiescence in the local plasma conditions close to the X-point, it is now important to establish how robust this observation is to changes in plasma conditions and operation of the machine. To this end a survey has been conducted across data taken on MAST with a similar camera configuration to that employed in the previous section. Table 1 details the MAST shots that have been surveyed, alongside their confinement mode, their magnetic configuration, and several plasma parameters. The database listed in table 1 covers a significant range of parameter space for plasmas in MAST. Across this database the QXR is observed in all cases without exception. It should be noted that connected double null (CDN) cases are limited because the signal level detected by the camera in these cases drops significantly due to a reduction in the density in the divertor. However for the high density CDN case in the database, the QXR was identified. In all cases the QXR extends from the separatrix to approximately the $\psi_N = 1.02$ flux surface, though it has not been possible to check with significant accuracy whether any small systematic change to the width of the QXR in $\psi_N$ is present. In real space the width of the QXR may change as the magnetic geometry is altered due to a change in the plasma current, transitioning to H-mode, or going to a double null configuration. The full analysis presented in section 3.1 has been carried out for each shot in the database. For brevity however the results will not be presented here. As a way of comparing the different shots the cross-correlation along a horizontal line at a vertical position that is one pixel below the X-point (in order to avoid including the selected pixel in the analysis) is shown in figure 8 for a subset of shots in table 1. The five individual graphs represent the five different positions of the selected pixel as it is moved radially away from the X-point, as in figure 5. Also shown is the cross-correlation from synthetic data-seta a) (solid line) and b) (broken line) and the position of the $\psi_N = 1.02$ flux surface in shot 29611 which characterises the extent of the QXR.

**Table 1.** Survey of plasmas from MAST upon which the QXR analysis detailed in section 3.1 has been carried out.

| Shot | Mode | Configuration | $\overline{n_e}(10^{19}m^{-2})$ | $I_p(MA)$ | $B_{tor}(T)$ | NBI($MW$) |
|------|------|---------------|---------------------------------|-----------|--------------|-----------|
| 29611 | L-mode | LSN | 10.8 | 0.62 | 0.58 | 0.0 |
| 29606 | L-mode | LSN | 8.8 | 0.63 | 0.59 | 0.0 |
| 29668 | L-mode | LSN | 11.5 | 0.62 | 0.59 | 0.6 |
| 29628 | L-mode | LSN | 11.7 | 0.63 | 0.57 | 1.2 |
| 29628 | H-mode | LSN | 17.8 | 0.62 | 0.52 | 1.2 |
| 29652 | L-mode | LSN | 9.9 | 0.63 | 0.57 | 1.2 |
| 29717 | L-mode | LSN | 11.3 | 0.63 | 0.57 | 1.6 |
| 29717 | H-mode | LSN | 18.6 | 0.63 | 0.53 | 1.6 |
| 29641 | L-mode | LSN | 11.2 | 0.42 | 0.52 | 1.2 |
| 29496 | L-mode | LSN | 11.4 | 0.42 | 0.51 | 3.2 |
| 29724 | L-mode | LSN | 11.3 | 0.78 | 0.62 | 1.6 |
| 29609 | L-mode | CDN | 11.8 | 0.62 | 0.59 | 0.0 |

As the selected pixel moves away from the X-point, the cross-correlation becomes more peaked as the structure that it adopts starts to represent the perpendicular shape of a filament rather than the parallel shape. It is notable that the deviation between synthetic data-set a) and the experimental data grows in the QXR with the synthetic data-set a) maintaining its peakedness due to the presence of filament cross-sections in the QXR which are not present in the experimental data. Synthetic data-set b) on the other hand compares favourably with the experimental database. Whilst the QXR has been identified in the H-mode cases, CDN case and cases with different $I_p$, these are excluded from figure 8 because the position of the X-point and geometry of the divertor leg varies, making direct comparison in the manner shown in figure 8 less clear. Carrying out the full set of analysis techniques employed in section 3.1 on any of the shots listed in table 1 shows that the QXR exists which suggests that the QXR is largely ubiquitous to MAST plasmas. Furthermore the QXR is approximately bounded by the $\psi_N = 1.02$ in each case, indicating that it is strongly related to the magnetic field structure.

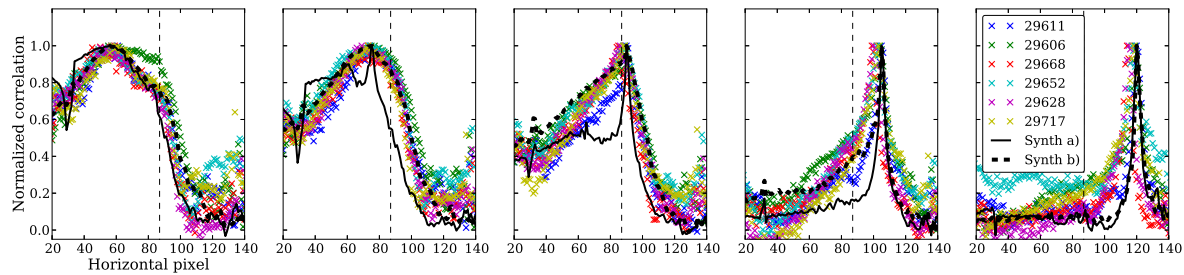

**Figure 8.** Cross-correlation calculated along a horizontal line displaced by one pixel below the X-point location (see fig 5) for a subset of shots listed in table 1. Also shown is the cross-correlation in synthetic data-seta a) (solid line) and b) (broken line). The vertical line shows the position of the $\psi_N = 1.02$ from shot 29611 which represents approximately the extent of the QXR. The five plots represent the same five positions of the selected pixel in figure 5.

## 5. Relationship between the QXR and target profiles

Figure 9 shows the heat flux measured by IR thermography as a function of major radius at the divertor target for shot 29611.
Highlighted on figure 9 is the QXR mapped to the divertor target. The QXR contains approximately 60% of the integrated heat flux shown which is close to the heat flux contained in one e-folding length of the profile. This shows that the role of the QXR on target profiles is non-negligible and warrants a closer analysis of the nature of transport within the QXR.
Previous analysis of Langmuir probe data at the outer divertor target on MAST [20] has shown a change in the standard-deviation of the ion-saturation current in a region that is consistent with the QXR. However the standard deviation does not reduce but increases towards the separatrix indicating that fluctuations exist throughout the SOL at the divertor target, contrary to the observations at the X-point. These near-separatrix fluctuations at the target are likely to be related to divertor-localised filaments as identified by Harrison et al on MAST [19] and previously by Maqueda et al on NSTX [21]. These filaments appear close to the separatrix, within the QXR, and are strongly aligned

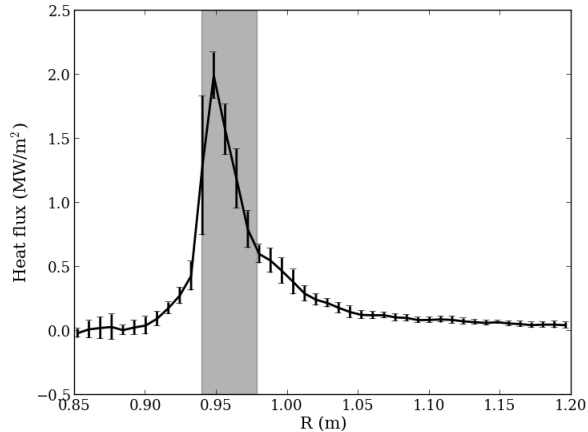

**Figure 9.** Heat flux at the outer divertor of MAST in shot 29611. The QXR mapped to the divertor target is shown as a shaded region and contains 60% of the total heat flux to the target.

to the magnetic field in the divertor leg. They have a small cross-field dimension which is inconsistent with the effects of shear and flux expansion around the X-point acting on a flux-tube from upstream, suggesting that such filaments are born locally within the divertor. Furthermore their localisation close to the separatrix ensures that they are born within the region of the divertor leg mapped out by the QXR. The structure of these divertor filaments can be observed in the background subtracted camera image in figure 2, however to elucidate this structure more clearly cross-correlation of a pixel in the camera image that is strongly affected by divertor filaments has been performed. This is compared to the filament structure in the outer SOL. Furthermore a time-delay (in units of frame count which are equivalent to $8\mu s$ in time) has been applied to the cross-correlation to track the evolution of the filaments across 5 frames of the movie. This is shown in figure 10

Across a centered window of five frames of the movie, the outer SOL filaments show no appreciable change in structure and do not show a drop in the correlation level indicating that their dwell time is typically longer than $16\mu s$ (twice the inter-frame time). The divertor localised filaments on the other hand show a clear reduction in the correlation level across two frames in the past or future, indicating that their typical dwell time is much shorter.

An ideal cross-diagnostic comparison would be an analysis of time-series data from the ion-saturation current, $I_{sat}$, measured on probes at the divertor target. Typically, however, the divertor Langmuir probes on MAST are run with a swept voltage to collect I-V characteristics. This sweeping makes it impossible to generate statistically suitable time-series of $I_{sat}$ in the shots analysed within this paper, so direct comparison is not possible. Luckily a small historic set of shots exists where the LPs were set to remain in ion-acceptance mode for the entirety of the shot, such that $I_{sat}$ time-series can be collected with decent statistics. Within this database, shot 16161 has been identified as comparable to the shots examined here. Figure 11 shows time-traces from a set of

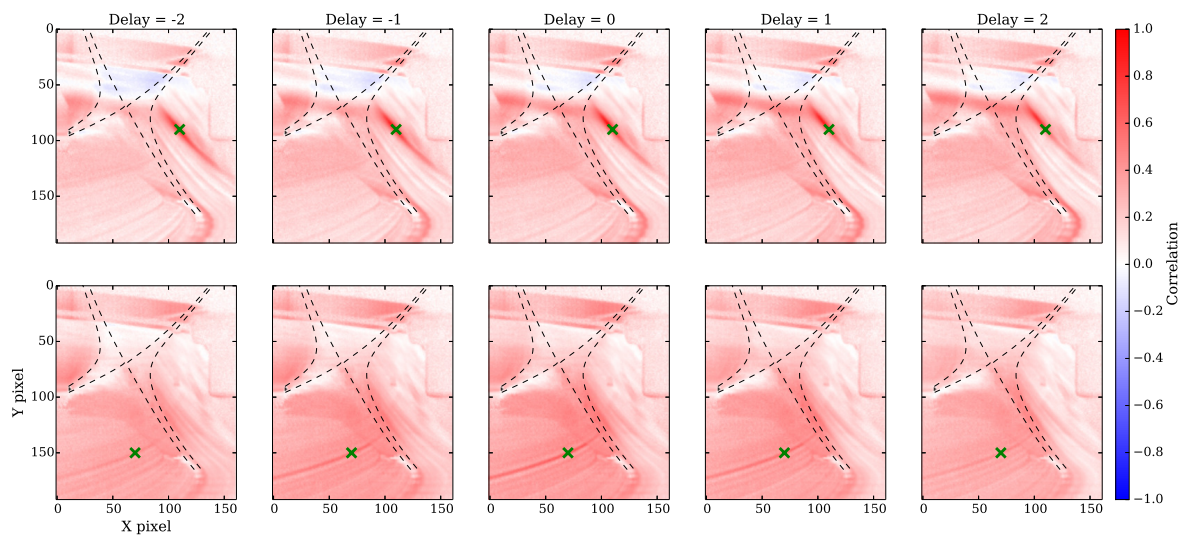

**Figure 10.** Cross-correlation showing the structure of filaments in the outer SOL (upper row) and filaments in the near SOL (lower row) with a delay introduced in units of frame count to show the typical evolution of such filaments.

5 radially separated LPs that span the SOL region, alongside a measurement of the auto-correlation time and power spectrum from the $I_{sat}$ time-series.

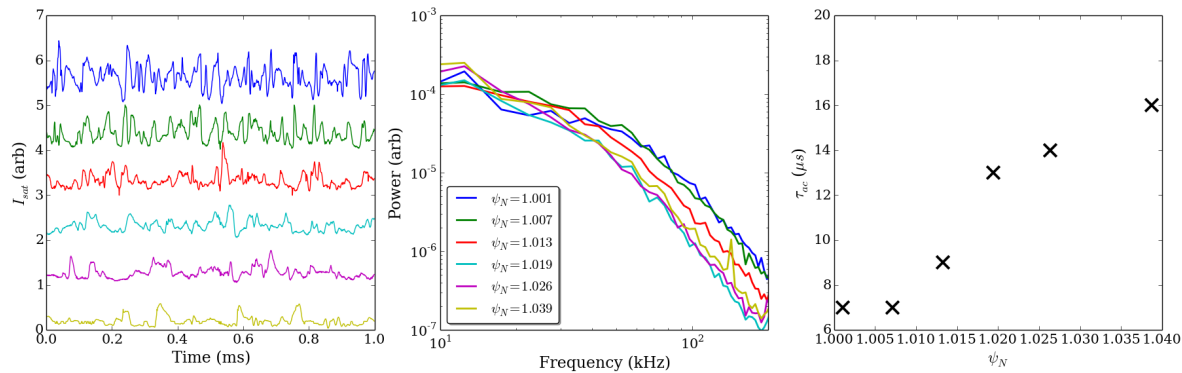

**Figure 11.** Analysis of $I_{sat}$ time-series from the divertor target of MAST shot 16161 from radially separated LPs. Left: 1ms of data from each probe. Note that artificial offsets have been applied to the data to allow for each dataset to be visualised on the same axis. Center: Power spectra from each probe at different radial positions. Right: Autocorrelation time, measured as the e-folding time of the autocorrelation function, as a function of $\psi_N$ at the divertor target.

The analysis of $I_{sat}$ at the divertor target shows a tendency towards higher frequency, shorter living events close to the separatrix and longer living, less frequent events away from the separatrix. Note that the precise separatrix location is subject to some uncertainty inherent in the EFIT reconstruction, however the radial variation of the autocorrelation time is consistent with the shorter lifetimes of divertor localised filaments as shown in figure 10 from the camera measurements. As already noted, it is not possible to make this comparison simultaneously with current data from MAST, however this would be an excellent avenue to pursue in the future.

## 6. Discussion

The results presented in this paper provide a complex multi-region picture of intermittent cross-field transport in the divertor volume. In the outer SOL filaments are able to connect from upstream all the way through to the divertor target. In the near SOL the X-point inhibits this connection, blocking filaments from upstream and instead allowing for the emergence of divertor localised filaments. These are quicker to dissipate than upstream filaments and may not be able to exist in regions where upstream filaments dominate, but can be driven in the quiescent region shielded by the X-point. In the PFR filaments are generated on the inner divertor leg and connect along magnetic field-lines to the outer target [19, 20]. Figure 12 attempts to present this complex zoology of filamentary structures schematically.

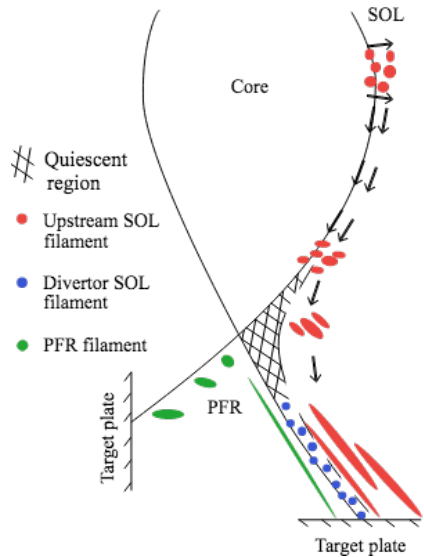

**Figure 12.** Schematic illustration of the different contributions to intermittent cross-field transport in the divertor volume observed on MAST.

There is evidence for this kind of complex multi-region transport on other devices. On NSTX a decorrelation between upstream and divertor images was detected close to the separatrix, but not in the far SOL [21]. Also noted was that the filaments observed at the divertor target close to the separatrix were of a higher frequency than those in the far SOL, consistent with the observations made here. A similar increase in frequency towards the separatrix was observed in JET in the ion saturation current of a Langmuir probe during a strike point sweep [31]. Such a change in filamentary characteristics close to the separatrix at the divertor therefore appear to be consistent between several devices. At the X-point, Tanaka *et al* [32] report a reduction in the fluctuation level of the ion-saturation current measured on a reciprocating probe as it is plunged towards (though slightly below) the X-point of JT-60U. This occurs despite very similar mean profiles, indicating weaker fluctuations in the vicinity of the X-point as observed in this paper.

The question now naturally arrises as to the physical mechanism causing the QXR. Whilst there is clearly significant uncertainty as to the cause of the QXR, which can not be ascertained from the imaging data shown here, there are several possibilities. Firstly, and most trivially, if filaments upstream are born only *outside* $\psi_N = 1.02$, the QXR will be present in the divertor as highlighted by figures 7 and 8 where a synthetic data set which excludes filaments from the QXR captures the experimental measurements. Given that there is currently no consensus on the origin of filaments, this is possible, however measurements of filament properties on NSTX suggest that it is unlikely [29]. Across a large database of measurements, Zweben *et al* [29] identify a significant proportion of filaments born within the separatrix which propagate radially into the SOL, suggesting other effects may be the cause of the QXR.

If filaments are born inside the separatrix, the timescale of their radial transport compared to their parallel transport may still naturally lead to a QXR region. This may be plausible, but would naturally lead to a graduated region of quiescence that should not conform to a flux surface. The observations suggest that immediately outside the QXR filaments are able to connect through to the divertor surface, whilst inside the QXR this does not appear to be the case.

Simple calculations of magnetic fieldline trajectories confirm that the sharp rise in magnetic shear close to the X-point significantly deforms the cross-section of flux-tubes [12]. This can impact the linear stability of the SOL, allowing for instabilities to be driven with eigenfunctions that are confined to being below the X-point [33]. This has also been shown theoretically to impact the motion of filaments by drastically raising cross-field gradients across the filament cross-section, leading to dissipation either through enhanced diffusion [17] or secondary instabilities such as the resistive drift-wave [34, 35]. At very large levels of magnetic shear the filament cross-section can drop to length scales below the gyro-scale, which may cause it to loose coherency [14]. In general it is unclear how a filament may act when only one of it's dimensions is sub-gyro scale (i.e a thin ribbon structure) however for circular filaments, FLR effects can be important in gyro-scale filaments [36]. Additionally strong shear flows are known to be driven in the vicinity of the separatrix due, at least in part, to the variation in the sheath potential [37]. This shear flow may interact strongly with filaments close the X-point, where the cross-section begins to distort, either causing the filament to break up or spread the filament poloidally thereby further reducing its coherency

Finally, it is well known that in any realistic plasma there is a degree of stochasticity to the magnetic field in the X-point region [38]. Filaments following magnetic field lines into such a stochastic region may loose coherency by virtue of their flux-tubes diverging in the stochastic field and their cross-sections loosing coherency. Such stochasticity may be induced by external error fields, however the plasma response to such fields would vary with plasma conditions, which is not a behaviour noted in the QXR. A good test of this effect may be to use small resonant magnetic perturbations to purposefully induce stochasticity around the X-point and observe how the structure of the QXR is varied. It is important to note that the lack of filaments in the QXR does not mean that filaments are unimportant close to the separatrix. On the contrary, if filaments cross the separatrix upstream then they must play a role in delivering heat and particles into the QXR, despite their loss in coherency. It then becomes important to establish the mechanism by which this coherency is lost in order to understand if future tokamaks will exhibit similar behaviour, or whether fluxes in the near separatrix will be more intermittent. There are also, as shown in section 5, fluctuations that contribute to the ion flux to the target that appear to be local to the divertor. It is now important to establish if these divertor filaments are related to upstream filaments, or are unrelated and intrinsic to the divertor leg. This is particularly relevant with a view to ITER and DEMO, where the spreading parameter $S$ of the typically used Wagner-Eich fitting function [39, 40], which represents the contribution of transport in the divertor to the target heat flux profile, can be the dominant parameter in the integral heat-flux width [41].

Finally, though the precise cause is unknown, the presence of the QXR indicates that the magnetic geometry of the divertor configuration can impact cross-field transport at and below the X-point. Alternative divertor concepts for DEMO such as the Super-X divertor [42, 43] and the snowflake divertor [44, 45] are being proposed and rely on changes to the magnetic topology of the divertor and, in the case of the snowflake, changes to the X-point region. Investigating the impact of such changes on the QXR and generally on fluctuations in the divertor region is therefore an important avenue for future study.

## 7. Conclusions

High speed imaging has been used in MAST to investigate the transient properties of the X-point region. Several measurement techniques have been employed to show that the region spanned by the separatrix and the $\psi_N = 1.02$ poloidal flux surface (here termed the Quiescent X-point Region or QXR) is devoid of filamentary structures and appears quiescent in the poloidal plane. OSM-EIRENE simulations rule out a local reduction in the neutral atomic density close to the separatrix as the cause of the observed quiescence. Furthermore careful comparison with a synthetically produced set of data shows that effects associated with the camera viewing geometry are not the cause. This leads to the conclusion that the quiescence observed is quiescence in the plasma conditions local to the X-point. Extension of the analysis techniques to a database of shots on MAST shows that the QXR is a ubiquitous feature in plasmas with different plasma parameters, geometry and confinement mode. When mapped to the divertor target the QXR contains approximately 60% of the integrated heat flux and shows higher frequency, shorter lived fluctuations in the ion saturation current compared to the outer SOL. This is consistent with the presence of short-lived divertor localised filaments observed in the near-separatrix region in the high-speed movies and suggests that such divertor-localised tranport may be having an impact on target profiles. This presents a multi-region view of cross-field transport in the divertor with the outer-SOL being dominated by filaments originating upstream that are blocked by the X-point in the inner SOL, allowing for the emergence of divertor localised filaments.

## 8. Acknowledgements

This work has been carried out within the framework of the EUROfusion Consortium and has received funding from the Euratom research and training programme 2014-2018 under grant agreement No 633053. This work has received funding from the RCUK Energy Programme [grant number EP/I501045]. The views and opinions expressed herein do not necessarily reflect those of the European Commission. To obtain further information on the data and models underlying this paper please contact PublicationsManager@ukaea.uk.

## 10. Appendix: Synthetic diagnostic

This appendix describes the synthetic diagnostic used to analyse filamentary structure in the divertor in section 3.1. The synthetic diagnostic treats an image as a linear combination of 'basis images'. Each basis image is created by projecting a magnetic field line trajectory onto the camera field of view (FOV) with the emission received by each pixel proportional to the length of fieldline within that pixel. The basis images are parameterised by their fieldline launch positions at the outboard midplane in major radius, $R$, and toroidal angle, $\phi$. A local plasma emissivity on a toroidal plane at the outboard midplane is then used as a weighting of the basis images to produce an image corresponding to the cameras view of that emissivity. This makes the assumption that the emissivity is constant along magnetic fieldline and that the neutral density is homogenous. Figure 13 shows 3 examples of randomly generated images in the case where filaments are seeding inside the QXR (upper row) and where they are not seeded inside the QXR (lower row).

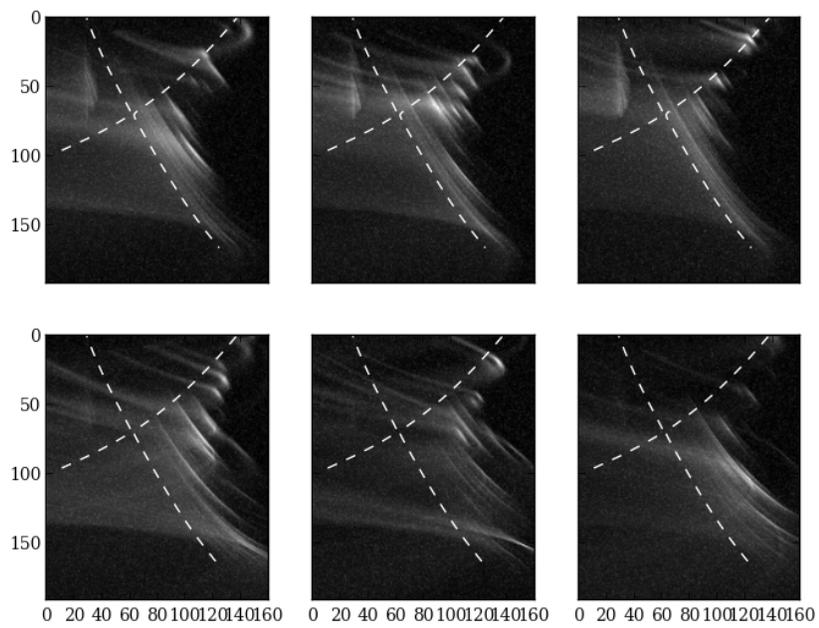

**Figure 13.** Example frames from the synthetic diagnostic with 5% salt noise added for cases where filaments are (upper) and are not (lower) seeded within the QXR